\documentstyle[prl,aps,epsf,floats]{revtex}

\begin{document}
\draft
\twocolumn[\hsize\textwidth\columnwidth\hsize\csname @twocolumnfalse\endcsname
\title{ \bf Quasi-degenerate self-trapping in One-Dimensional 
         Charge Transfer Exciton}
\author{
Andrei S. Mishchenko$^{1,2}$ and Naoto Nagaosa$^{1,3}$}
\address{
$^1$Correlated Electron Research Center,
Tsukuba 305-0046, Japan \\
$^2$RRC 'Kurchatov Institute', 123182, Moscow, Russia \\
$^3$Department of Applied Physics, The University of Tokyo, 7-3-1 Hongo, 
Bunkyo-ku, Tokyo 113, Japan}
\maketitle
\begin{abstract}
The self-trapping by the nondiagonal particle-phonon
interaction between two quasi-degenerate energy levels of excitonic 
system, is studied. We propose this is realized in 
charge transfer exciton, where the directions of the polarization
give the quasi-degeneracy. 
It is shown that this mechanism, unlike the conventional diagonal one, 
allows a coexistence and resonance of the free and self-trapped states 
even in one-dimensional systems and a quantitative theory for the optical 
properties (light absorption and time-resolved luminescence) of 
the resonating states is presented.  
This theory gives a consistent resolution for the long-standing 
puzzles in quasi-one-dimensional compound A-PMDA.
\end{abstract}
\vskip1pc]
\narrowtext

The charge transfer (CT) exciton in one-dimension (1D)
attracts great interests 
due to its peculiar features, e.g., 
strong interaction with the phonons, magnons, and large
nonlinearity in the optical responses.
However the theoretical understanding of it is rather limited compared 
with that of Wannier and Frenkel excitons,
because of its intermediate radius and strong coupling nature.
E.g., although the prototypical compound Anthracene-PMDA (A-PMDA),
which shows unusual vibronic structure in the optical spectra,
is thoroughly experimentally investigated from 70-th till nowdays 
\cite{HaPhMo75,BrPh80,Ha74,Ha77,ElWe85,KuGo94,KuGo97,KuGo99}, even
qualitative  theoretical understanding it's optical
response is missing.    
 
The most dramatic phenomenon which occurs at intermediate or strong coupling
with the phonons is the self-trapping  
where the free (F) state of polaron with small lattice deformation coexists
and resonates with the state with strong lattice 
relaxation \cite{R82,UKKTH86}.
Although the latter state is Bloch invariant it is traditionally classified
as being self-trapped (ST) to stress larger degree of ionic distortion
around the quasiparticle.   
However, this self-trapping scenario was not
considered yet since the F-ST resonance is prohibited by Rashba-Toyozawa 
theorem (RTT) in 1D systems 
\cite{R82,UKKTH86}.  
Although the RTT is probably valid for 
one quasiparticle state in the relevant energy range, the
neglect of it's domain of definition caused severe delusion in the 
physics of 1D systems because RTT was treated as the strict 
ban for F-ST coexistence in 1D. 
However, when there are two quasi-degenerate quasiparticle states, 
i.e. the conditions of the theorem are violated, one can consider 
the quasi-degenerate self-trapping (QDST) mechanism, when the F-ST
resonance is driven by nondiagonal interaction with phonons with 
respect to quasiparticle levels. An example of QDST was observed  
in the mixed valence systems \cite{KiMi}.
 
In this paper we first demonstrate by exact diagrammatic Monte Carlo method 
\cite{MM} that QDST is general phenomenon pertinent for 1D lattice, 
study the optic properties 
of resonating QDST states, and resolve long-standing puzzle of the
quasi-1D system A-PMDA 
\cite{HaPhMo75,BrPh80,Ha74,Ha77,ElWe85,KuGo94,KuGo97,KuGo99}.

{\it QDST existence in 1D by exact solution.} 
The quasi-1D compound A-PMDA consists of one-dimensional arrays of
alternating donor (D) and acceptor (A) molecules \cite{HaPhMo75,BrPh80}.
There are two possible configurations of CT exciton, i.e.,
$A^0 D^+ A^-$ and $A^- D^+ A^0$, and the    
symmetric $\phi_s(r)$ and antisymmetric $\phi_a(r)$ 
linear combination of these two constitutes the nearly degenerate
two branches of excitons. 

The minimal model to demonstrate F-ST coexistence in 1D
systems  
involves one optic phonon branch with frequency $\omega=0.1$, 
two quasiparticle branches with energies 
$\varepsilon_{1,2}(q) = \Delta_{1,2} + 2[1-cos(q)]$, 
where $\Delta_1=0$ and $\Delta_2=1$, and the quasiparticle-phonon coupling
\begin{equation}
\hat{H} = 
i \sum_{k,q} \sum_{i,j=1}^{2}  
\sqrt{\gamma_{ij}} \phi_{ij}(q) (b^{\dagger}_{q} - b_{-q}) 
c_{i,k-q}^{\dagger} c_{j,k} + h.c.
\label{1}
\end{equation}
Here $c_{j,k}$ and $b_{q}$ are annihilation operators 
for the quasiparticle of branch $j$ with momentum $k$ and for the phonon
with momentum $q$. One can restrict himself by one phonon mode since, even
in case of the different symmetry of excitons, the same phonon branch 
contributes both to the diagonal and off-diagonal coupling \cite{KM90} and 
drop, without affecting the conclusions, the $q$-dependence 
($\phi_{ij}(q) \equiv 1$) of the interaction for simplicity.
The energy $E_L$, the average number of phonons  
$\bar{N} \equiv \langle \Psi_L \mid \sum_q b^{\dagger}_{q} b_{q} 
\mid \Psi_L \rangle$, and Z-factors of the n-phonon states
$$
Z(n) \equiv \sum_{i=1}^{2} \sum_{q_1 ... q_n} 
\mid
\alpha_i(q_1,...,q_n) 
\mid^2
$$ 
in phonon cloud of lowest zero-momentum eigenstate  
$$
\Psi_L =  \sum_{i=1}^{2} \sum_{n=0}^{\infty} \sum_{q_1 ... q_n}
\alpha_i(q_1,...,q_n) c^{\dagger}_{i,-q_1...-q_n}
b^{\dagger}_{q_1} ... b^{\dagger}_{q_1} \mid \mbox{vac} \rangle 
$$
(Here $\mid \mbox{vac} \rangle$ is the vacuum state.)
were calculated by the exact diagrammatic Monte Carlo technique 
\cite{MM} which was generalized in the present paper
to the case of two-quasiparticle bands 
with interaction Hamiltonian (\ref{1}).
\begin{figure}[th]
\epsfxsize=0.45\textwidth
\epsfbox{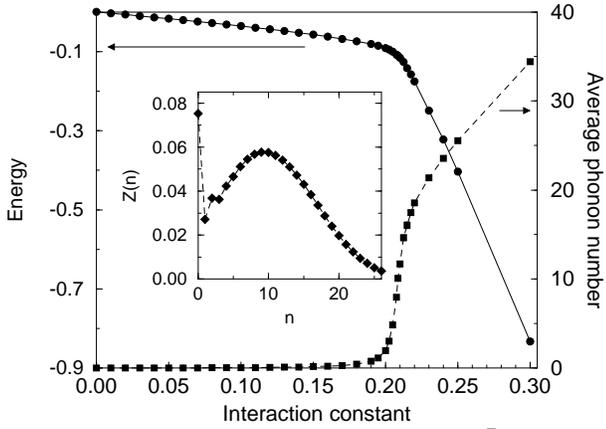}
\caption{Energy (solid line, circles) and $\bar{N}$ (dashed line, squares)
dependence on $\gamma_{12}$ at $\gamma_{11}=0$ and $\gamma_{11}=0.05$. 
Inset: Z-factors distribution for 
$\gamma_{12}=0.2087$. The statistic errorbars are much smaller than the 
point size.}
\label{fig:fig1}
\end{figure}
In the vicinity of critical $\gamma_{12}^c \approx 0.2087$ the dependence of 
the energy $E_L$ on $\gamma_{12}$ rapidly changes, and the average number of 
phonons $\bar{N}$ demonstrates sharp crossover to higher values 
(Fig.~\ref{fig:fig1}). 
Hence, according to \cite{R82,UKKTH86}, the 
F-state, which is lowest for small values of nondiagonal interaction 
constant,  resonates with the ST-state (see inset in Fig.~\ref{fig:fig1}) at 
$\gamma_{12}^c$ and 
ST-state is the lowest one for large values of $\gamma_{12}$ \cite{0}. 
The system indeed undergoes 
crossover around the critical constant rather than a phase 
transition due to mixing of resonating eigenstates \cite{GeLo91}.
 
\begin{figure}[th]
\epsfxsize=0.4\textwidth
\begin{center}
\epsfbox{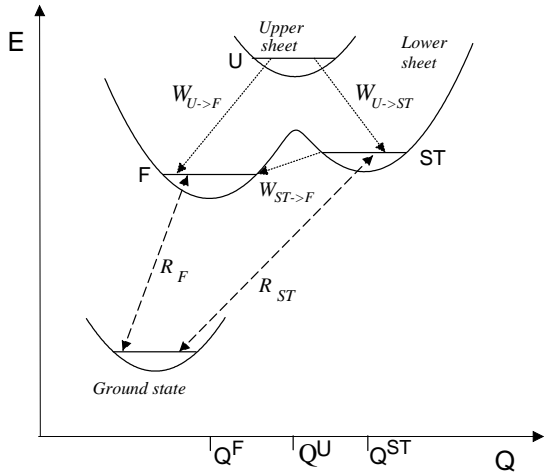}
\end{center}
\caption{Schematic configurational diagram. The dotted arrows correspond to 
nonradiative processes, and the dashed arrows indicate the radiative 
(absorption/emission) transitions.} 
\label{fig:fig2}
\end{figure}

{\it Strong-coupling limit model.}
If the quasiparticle is exciton, one can study the optical transition 
between the ground state $G$ and the  exciton-polaron states $\Psi_L$
with zero momentum (see Fig.~\ref{fig:fig2}).
To study the optical properties we consider the case when the coexistence 
occurs in the strong coupling limit, i.e. when 
$\gamma_{12}^2 / \omega \gg \Delta$ and, therefore, the adiabatic 
approximation is justified \cite{Ber83}. 
In this case one can find for given lattice deformation 
$Q=\{Q_1, ..., Q_N\}$ (N is the number of lattice modes) 
the adiabatic potential relieves 
$\epsilon_{\Lambda}(Q)$ 
($\Lambda$ is the index of lower $l$ and upper $u$ sheet) and 
express the eigenfunctions
$\psi_{\Lambda}({\bf r},Q) = 
\alpha^{\Lambda}_{1} (Q) \phi_1({\bf r}) + 
\alpha^{\Lambda}_{2} (Q) \phi_2({\bf r}) $
in terms of the excitonic wave functions $\phi_{1,2}({\bf r})$
for undistorted lattice \cite{KiMi,Ber83}.
In case of F-ST coexistence the lower sheet of the adiabatic potential 
contains two minima at different lattice distortions 
$Q^{t}$ ($t=F,ST$) which correspond to states with different 
lattice relaxation energy.  
When the minima are separated by large potential barrier one can neglect
the $Q$-dependence of the electronic functions and
introduce vibrational states
$\chi_{t}^{\{f\}} (Q - Q^t) =
\prod_{n=1}^N 
\chi_{t}^{f_n} (Q_n - Q_n^t)$
with quantum numbers 
$\{f_n\}$.
The low laying eigenstates of the adiabatic approximation are 
\begin{equation}
\tilde{\psi}_{t}^{\{f\}}({\bf r},Q) = 
\psi_{l} ({\bf r}, Q^t) \chi_{t}^{\{f\}} (Q - Q^t);
\; \; t=F,ST,
\label{2} 
\end{equation}
with the eigenvalues 
$E_{t}^{\{f\}} = 
\epsilon_{l}(Q^t) + \sum_{n=1}^{N} \omega_t^{n}(f_n+1/2)$.
The adiabatic basis (\ref{2}) is an approximate one due to 
nondiagonal matrix elements of nonadiabatic operator \cite{KiMi}. 
E.g., at critical $\gamma_{12}^c$
the ground state is already a mixture of weakly/strongly deformed states 
(inset in Fig.~\ref{fig:fig1}). 
However, the basis (\ref{2}) is valid
when interaction constant is 
close enough to the critical value $\gamma_{12}^c$ to create two 
wells but far enough to neglect the mixing of F and ST
sets.
 
{\it Spectrum of optical absorption.}
The optical properties of A-PMDA give the evidence of
the strong exciton-phonon interaction \cite{Ha74}. 
The low energy part of the absorption spectrum consists of the 
zero phonon line (ZPL) and a broad band 
with the full width at half maximum (FWHM) of around \cite{Ha77,BrPh80}
500 cm$^{-1}$. 
Therefore, the optical response qualitatively remind lineshape
of F-color centers absorption spectra which correspond to transitions between 
displaced oscillators \cite{HR50,OR53}. 
However, all attempts to interpret the optical properties 
of A-PMDA \cite{HaPhMo75,BrPh80,Ha74,Ha77} encountered 
apparent contradictions with the model \cite{HR50,OR53}: 
(i) absorption and emission spectra can not be described as 
one series of the equidistant lines \cite{BrPh80}; 
(ii) the value of Huang-Rhys parameter $S$, evaluated 
from the ratio of ZPL and total oscillator strength, 
is inconsistent with the intensity distribution of the broad
band [E.g., the value $S=4$ extracted from 
the oscillator strengths of ZPL leads to the estimate FWHM=140cm$^{-1}$.
\cite{BrPh80}.].

These long-standing puzzles can be interpreted (such treatment has never 
been attempted due to the RTT) as response of the 
coexisting  F and ST states.
To calculate induced
cross section $\sigma^{\mbox{\scriptsize abs}}(E)$ at energy $E$ of optical 
absorption which corresponds to transitions from the set 
$\psi_{G}^{\{\alpha\}} ({\bf r},Q) = 
\phi_G({\bf r}) \chi_{G}^{\{\alpha\}}(Q)$
of $\chi_{G}^{\{\alpha\}}(Q)$ 
phonon states of ground electronic state $\phi_G({\bf r})$ 
to the set of excited states in (\ref{2}) one has to average over initial 
states $\{\alpha\}$ and sum over final states in (\ref{2}). 
Using the standard approach \cite{Dex58}, 
defining the electronic dipole matrix elements
${\bf M}_i = \int d{\bf r} ( \phi_G({\bf r}) )^* {\bf r} \phi_i({\bf r})$, 
and introducing an average optic phonon frequency $\bar{\omega}_t$ one 
can take advantage of Huang-Rhys method \cite{HR50}
and obtain for zero temperature
\begin{equation}
\sigma^{\mbox{\scriptsize abs}}(E) = 4\pi^2e^2/(3c) \langle E \rangle 
\sum_{t=F,ST} {\cal M}_t 
F_t^{\mbox{\scriptsize abs}}(E),
\label{4}
\end{equation}
where
\begin{equation}
F_t^{\mbox{\scriptsize abs}}(E) = 
\frac{e^{-S_t}}{\bar{\omega}_t}
\sum_{p=0}^{\infty}
\frac{S_t^p}{p!}
{\cal S} \left( E - \epsilon_l(Q^t) - p\bar{\omega}_t,
\Upsilon_t^p \right)
\label{5}
\end{equation}
(Here $S_t=(2N)^{-1}\bar{\omega}_t \sum_{n=1}^{N}(Q^t_n)^2$ is the 
Huang-Rhys factor.)
is the normalized shape function, and 
\begin{equation}
{\cal M}_t = \sum_{i,j} 
\left|
\left( {\bf M}_i \alpha_i^l(Q^t) \right)^*
{\bf M}_j \alpha_j^l(\tilde{Q^t})
\right|
\label{6}
\end{equation}
is the weighting coefficient. Here $e$ and $c$ is 
electron charge and light velocity, $\langle E \rangle$ is the average 
energy of transitions,  
${\cal S}(x,\Upsilon_t^p) = (\Upsilon_t^p/2\pi)
(x^2+(\Upsilon_t^p/2)^2)^{-1}$, and
$\Upsilon_t^p$ is the spectroscopic linewidth.
We note that for $T \ne 0$ one can also 
use approach \cite{HR50} since the average is performed over the one well 
of the ground state \cite{1}. 

The contradiction (i) in the optical absorption spectrum of A-PMDA
is resolved when one note that experimental data
\cite{BrPh80} contain two series of
satellites in absorption spectrum: one arises from the
ZPL with the step $\bar{\omega}_{F}\sim$27 cm$^{-1}$ (0 - 29 - 55 - 83 -
111 cm$^{-1}$) and the second one (13 - 29 - 47 - 62 - 77
- 88 cm$^{-1}$) with the overtone $\bar{\omega}_{ST}\sim 12$ cm$^{-1}$
originates from the 13 cm$^{-1}$ satellite.
The contradiction (ii) gains natural explanation since 
the spectrum is the superposition of
the two contributions (\ref{4}) from  
F and ST states with different parameters $S_F$ and $S_{ST}$ 
\begin{figure}[th]
\epsfxsize=0.40\textwidth
\epsfbox{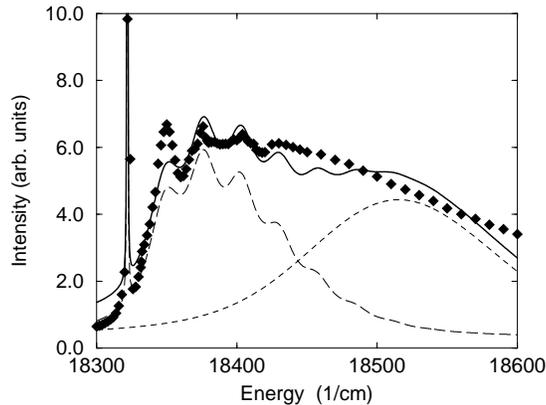}
\caption{Comparison of light absorption experimental data (diamonds) with 
model curve (solid line) consisting of optic response of coexisting $F$ 
(dashed line) and $ST$ (dotted line) states with $S_F=2.5$, 
$S_{ST}=13.3$, and ${\cal M}_F/{\cal M}_{ST}=0.82$. 
The onsets are $\epsilon_{F,ST}(Q^{F,ST})=18322\!/\!18334$ cm$^{-1}$  
and linewidthes are $\Upsilon_{F,ST} = 29\!/\!72$ cm$^{-1}$ 
($\Upsilon_{F}^{p=0}=2.5$ cm$^{-1}$).} 
\label{fig:fig3}
\end{figure}
(Fig.~\ref{fig:fig3}). 

{\it Luminescence from QDST states}.
The main puzzle in luminescence spectra of A-PMDA is that
although the pattern of satellites is symmetrically repeated in 
absorption and luminescence spectra, the intensity distribution of phonon
sidebands is significantly different \cite{UKKTH86}. 

The experimental conditions of time-resolved luminescence spectroscopy 
are equivalent to the situation when at a
moment $t=0$ the system is excited to the upper sheet $U$  whereas
F and ST wells are empty.
The rates of the intrawell nonradiative
processes are significantly larger than that of 
interwell ones $w_{U \leftrightarrow ST}$, $w_{U \leftrightarrow F}$
and $w_{ST \leftrightarrow F}$ (see Fig.~\ref{fig:fig2}), 
because successive relaxation of the system within one well does not demand 
many-phonon transitions whereas those are compulsory for the interwell 
jumps. Due to this first link of the "bottlenecks hierarchy" (BH), 
the states in one well can be treated as being in 
quasi-thermal equilibrium and only summary
populations $P_U(t)$, $P_{ST}(t)$ and 
$P_F(t)$ have to be considered as dynamical variables with initial conditions 
$P_U(t=0)=1$ and $P_{ST}(t=0)=P_F(t=0)=0$. 
The differential intensities $I^{U,ST,F}(t)$
of luminescence can be expressed in terms of radiative rates 
$R_U$, $R_{ST}$, and $R_F$ as 
$I^{U,ST,F}(t) = R_{U,ST,F} P_{U,ST,F}(t)$, and non time-resolved experimental
setup is defined by total yields  
$Y^{U,ST,F}_{\mbox{\scriptsize int}} = 
R_{U,ST,F} \int_0^{\infty} P_{U,ST,F}(t)$.
If the lattice distortion $Q^{U}$ in the upper sheet minimum and that for
$ST$-minimum of the lower sheet almost coincide whereas the lattice 
relaxation $Q^{F}$ for the minimum $F$ differs 
significantly (such situation can be easily obtained in strong-coupling limit
model), there is the second link of the BH
$w_{U \leftrightarrow ST} \gg 
\max \left\{ w_{U \leftrightarrow F}, w_{ST \leftrightarrow F} \right\}$,
which is due to Franck-Condon suppression of nonradiative 
processes $U \leftrightarrow F$ and $ST \leftrightarrow F$. 
At low temperatures (when, e.g., ST state is metastable and the rates  
$w_{U \leftarrow ST}$, 
$w_{U \leftarrow F}$, and $w_{ST \leftarrow F}$ are negligibly small)
the dynamics of the populations is reduced 
to the system of equations \cite{2}
\begin{eqnarray}
\dot{P}_U(t)  & = & 
- ( w_{U \to ST} + w_{U \to F} ) P_U(t)
\nonumber \\
\dot{P}_{ST}(t)  & = & 
- ( w_{ST \to F} + R_{ST} ) P_{ST}(t) + w_{U \to ST} P_U(t)
\nonumber \\
\dot{P}_F(t)  & = & 
- R_F P_F(t) + w_{U \to F} P_U(t) + w_{ST \to F} P_{ST}(t). 
\nonumber
\end{eqnarray}

The differential intensities of the luminescence from the states 
$T$ and $F$ are 
\begin{eqnarray}
I_{ST}(t) & = & 
\frac{R_{ST} w_{U \to ST}}{R_{ST}+w_{ST \to F}-w_{U \to ST}-w_{U \to F}}
\times
\nonumber
\\
&&
\left[
e^{ - (w_{U \to ST} + w_{U \to F} ) t } -
e^{ - (w_{ST \to F} + R_{ST}) t }
\right]
\label{10}
\end{eqnarray}
and
\begin{eqnarray}
I_F(t) =  
R_F  &&
\left\{ 
\frac{(w_{U \to F}+{\cal D})}{(R_F-w_{U \to ST}-w_{U \to F})} \times
\right.
\nonumber
\\
&&
\left[
e^{ - (w_{U \to ST} + w_{U \to F}) t} -
e^{ - R_F t}
\right] + 
\nonumber
\\
&&
\frac{{\cal D}}{(R_F-R_{ST}-w_{ST \to F})} \times
\nonumber
\\
&&
\left.
\left[
e^{ -  R_F t } -
e^{ - (R_{ST} + w_{ST \to F} ) t } 
\right]
\right\},
\label{11}
\end{eqnarray}
respectively. Here 
${\cal D} = 
w_{U \to ST} w_{ST \to F}/(R_{ST}+w_{ST \to F}-w_{U \to ST}-w_{U \to F})$.
Introducing the dimensionless quantities 
${\cal V}_{U \to F, U \to ST} =
w_{U \to F, U \to ST}/(w_{U \to ST}+w_{U \to F})$
and $\eta_{ST} = R_{ST} / w_{ST \to F}$
one obtain  the integral yields as
$
Y_{ST} = \eta_{ST} {\cal V}_{U \to ST}(1+\eta_{ST})^{-1}
$
and
$
Y_F = (1+\eta_{ST} {\cal V}_{U \to F})(1+\eta_{ST})^{-1}
$, 
respectively.
There are some domains of model parameters when one can observe vivid 
qualitative features.
E.g., second link of the BH leads to relation $Y_{ST} / Y_F \gg 1$
provided the inequality $R_{ST}>w_{ST \to F}$ takes place. 
Therefore, due to specific BH of the relaxation
processes it is possible that, e.g., the $F$ state is
distinctly observed in the light absorption spectra but is almost not seen 
in the spectra of non time-resolved luminescence. 
At the same time if criterion $R_F<R_{ST}$ is satisfied, 
at large times $t > R_F^{-1}$ the ratio of differential intensities 
$I_{ST}(t) / I_F(t) \ll 1$ is opposite. Note, that this relation implies 
that ${\cal M}_{ST} > {\cal M}_F$ because the rate of
radiative processes is governed by the matrix elements (\ref{6}).
Due to the first link of the BH at $T=0$ only ground vibrational states of 
the wells $ST$ and $F$ are populated at a moment of radiative transition. 
Therefore, the many-well structure leaves intact 
Huang-Rhys approach since no average over the initial states is 
necessary. Hence, the differential spectrum of the luminescence is
$
\sigma(E,t) = 
I_{ST}(t) F_{ST}^{\mbox{\scriptsize em}}(E) +
I_F(t) F_F^{\mbox{\scriptsize em}}(E)
$.
Here the emission normalized shape function 
$F_t^{\mbox{\scriptsize em}}(E)$
can be obtained from (\ref{5}) by replacing
$p \to -p$ in 
the function ${\cal S}$. For the lineshape of non time-resolved
experiment one has to use integral yields $Y_{ST}$ and 
$Y_{F}$ instead of 
differential intensities (\ref{10}) and (\ref{11}). Note, that parameters 
$S_F$ and $S_{ST}$ are the same as in absorption experiment.

Finally, the difference between absorption and luminescence spectra
pattern arises in A-PMDA due to the  
BH which manifests itself in the mirror image of sidebands  
coming from ST and F series in absorption and luminescence spectra, which
relative intensities are different in two types of experiment. 
Moreover, due to the BH both 13 cm$^{-1}$ peak \cite{Ha77} and the whole 
broad structure \cite{Ha74,UKKTH86} have inverse relative
spectral weights in luminescence and absorption spectrum.   
More evidences can be found in experiments on time-resolved luminescence
because proper fit (Fig.~\ref{fig:fig3}) of the experimental curve 
\cite{BrPh80} demands ${\cal M}_F/{\cal M}_T<1$ in (\ref{4}), and, hence,  
criterion $R_F<R_{ST}$ is satisfied.

{\it Conclusions.}
We have demonstrated that the coexistence of the 
F and ST states is possible even in 1D lattices, 
owing to the possibility of the QDST mechanism. 
When there are more than one electronic levels of a 
particle 
within the energy range of the lattice relaxation energy, 
F and ST states can coexist and even resonate. 
Hence, although the Rashba-Toyozawa theorem is not overthrown and
remains valid within its domain of definition, there is the possibility 
of the coexisting of ST and F states in 1D systems. 
We have found several features which are unique for QDST states and
can be checked in optical experiments.
We have shown, that the QDST mechanism 
provides comprehensive description of the long-standing mystery of the 
optical properties of the quasi-1D compound A-PMDA. 

We acknowledge  M.\ Gonokami, 
B.\ V.\ Svistunov, and Y.\ Tokura for critical 
discussions. This work was supported by Priority Areas Grants 
and Grant-in-Aid for COE research from the Ministry of Education, 
Science, Culture and Sports of Japan, and RFBR 01-02-16508.


\begin{references}
\bibitem{0} Our calculations for two- and three-dimensional
cases show that the QDST phenomenon is possible for all dimensionalities.
\bibitem{1}
Method \cite{HR50} is valid when the difference 
between phonon frequencies in
F and ST states can be neglected. 
Otherwise expressions in ref.\ \cite{OR53}, which do not alter 
the gross features of (\ref{4}), are desirable for better fit.
\end{references}
\end{document}